# Enhancing the dielectric, electrocaloric and energy storage properties of lead-free $Ba_{0.85}Ca_{0.15}Zr_{0.1}Ti_{0.9}O_3$ ceramics prepared via sol-gel process

H. Mezzourh [1,2], S. Belkhadir [1], D. Mezzane [1], M. Amjoud [1], E. Choukri [1], A. Lahmar [2], Y. Gagou [2], Z. Kutnjak [3], M. El Marssi [2]

[1] IMED, Cadi-Ayyad University, Faculty of Sciences and Technology, BP 549, Marrakech, Morocco.

[2] Laboratory of Physics of Condensed Matter (LPMC), University of Picardie Jules Verne, Scientific Pole, 33 rue Saint-Leu, 80039 Amiens Cedex 1, France.

[3] Jozef Stefan Institute, Jamova Cesta 39, 1000 Ljubljana, Slovenia

**Abstract**

The $Ba_{0.85}Ca_{0.15}Zr_{0.1}Ti_{0.9}O_3$ (BCZT) ceramics were successfully prepared by sol-gel process, and sintered at 1420°C. The effect of sintering times (2, 4 and 6 h) on structural, microstructural, electric properties, energy storage, and electrocaloric effect was systematically examined. X-ray diffraction (XRD) results show that all the samples crystallize in the pure perovskite structure. The morphotropic phase transition from the tetragonal to the orthorhombic phase (T-O) was identified and confirmed by Rietveld refinement. The BCZT ceramic sintered at 1420°C for 4h possesses a good relative density of 98%, and exhibits optimal properties with a high dielectric permittivity ($\varepsilon_{r,max}$) ~16310, a large electrocaloric effect coefficient ($\xi$) ~ 0.244Kmm/kV and an energy density storage of ~ 10.61 mJ/cm$^3$.

## 1. Introduction

Since many years, lead-based piezoelectric materials such as lead zirconium titanate $PbTiO_3$–$PbZrO_3$ (PZT) **[1]**, $Pb(MgNb)O_3$–$PbTiO_3$(PMN-PT), and $Pb(Ni_{1/3}Nb_{2/3})O_3$–$PbHfO_3$–$PbTiO_3$ (PNN–PHT) **[2]** are extensively used in various functional applications such as non-volatile memory, sensors, actuators, transducers, capacitors, generators and energy harvesting devices, **[3]–[6]** due to their eminent electrical performance close to the morphotropic phase boundaries. However, these compounds have the undeniable disadvantage of toxicity and volatility of lead. The use of these materials has contributed in a serious environmental and human health problems **[7], [8]**. Recently, many scientific teams have been searching for new eco-friendly alternative materials with advanced properties. By investigating various lead-free materials with perovskite structure ($ABO_3$) such as $BaTiO_3$(BT), $(Bi_{0.5}Na_{0.5})TiO_3$-$BaTiO_3$(BNT-BT), $(K_{0.5}Na_{0.5})NbO_3$ (KNN) and $Ba_{1-x}Ca_xTi_{1-y}Zr_yO_3$ (BCZT) **[9]–[11]**. Among of all the above materials, barium calcium zirconate titanate ($(Ba,Ca)(Zr,Ti)O_3$, BCZT) exhibits a super large piezoelectric coefficient ($d_{33}$) up to 620 pC/N **[12]** in a bulk ceramic with perovskite structure, and possesses a high dielectric permittivity value of ~ 17184 **[13]** as well as important ferroelectric properties and low loss tangent (tan δ). It is thought that the high-performance of BCZT ceramics may stem from the morphotropic phase boundary that is a region of specific interest for electroceramic materials **[14]**. It is well known that, the chemical and electrical properties of electroceramics depend strongly on the synthesis methods and preparation conditions. Several synthesis techniques were employed to prepare perovskite powders such as solid-state technique, hydro-solvothermal synthesis and sol-gel process **[15]–[18]**. Many papers have reported that the sol gel method provides a better homogeneity, chemical purity, and stoichiometric composition of the resultant phase comparing to the ceramics prepared by solid state method **[19]**. In general, the sintering factors such as temperature and dwelling time affect the structural and electrical properties of BCZT ceramics **[20], [21]**. Cai et al. reported that excellent ferroelectric properties (2Pr = 31.62 µC/cm$^2$, 2Ec = 3.46 kV/cm) of $Ba_{0.85}Ca_{0.15}Zr_{0.1}Ti_{0.9}O_3$ ceramics obtained by sol-gel method, can be achieved by sintered at 1500 °C for 10h **[22]**. Moreover, an interesting recoverable energy density value of 0.52 J/cm$^3$ reported in $Ba_{0.85}Ca_{0.15}Zr_{0.1}Ti_{0.9}O_3$ by Wang et al. **[19]** indicates that this ceramic could be a promising candidate for energy storage applications. On the other hand, Rui Liu et al. found that the ceramic sintered at 1420°C exhibits a good dielectric permittivity with a weak dielectric loss ~ 0.015 **[23], [24]**. In that regard, our group has reported that BCZT-spherical nanoparticles and BCZT-rod like ceramics elaborated by surfactant-assisted solvothermal route exhibit enhanced dielectric, ferroelectric and electrocaloric properties **[25]**. In addition, we have studied the influence of the sintering temperature on the microstructure, dielectric and diffusivity of the BCZT ceramic prepared by sol-gel method **[26]**. In the present work, we investigate the effect of sintering dwell time at 1420°C on the dielectric, electrocaloric and energy storage properties of BCZT ceramics prepared by sol-gel technique.

## 2. Experimental details

### 2.1. Chemicals

For the preparation of BCZT ceramics, the following materials purchased from Sigma-Aldrich and Alfa Aesar, were used: barium acetate $Ba(CH_3COO)_2$, calcium nitrate tetrahydrate $Ca(NO_3)_2 \cdot 4H_2O$, titanium (IV) isopropoxide $Ti[OCH(CH_3)_2]_4$ and zirconium (IV) oxychloride octahydrate $ZrOCl_2 \cdot 8H_2O$ as the starting materials, acetic acid and 2-Methoxyethanol were used as the solvents.

### 2.2. Synthesis procedure

The $Ba_{0.85}Ca_{0.15}Zr_{0.1}Ti_{0.9}O_3$ (BCZT) ceramic was synthesized by sol-gel route. Firstly, the $Ba(CH_3COO)_2$ and $Ca(NO_3)_2 \cdot 4H_2O$ were initially dissolved completely in acetic acid according to the stoichiometric ratio under continuous stirring. Afterward, stoichiometric amount of $ZrOCl_2 \cdot 8H_2O$ was dissolved in 2-Methoxyethanol solution, then $Ti[OCH(CH_3)_2]_4$ was added and stirred. In the next step, this solution was added to above (Ba, Ca) precursor solution and stirred for 1h at room temperature to yield a clear transparent BCZT solution. The pH was adjusted with ammonia to 6. The obtained sol was dried at 120°C for 24 h to eliminate the solvents, and then grounded to fine powder in a mortar. The powder was calcined for 4 h at 1000 °C in air to form a pure crystalline phase. Then, the calcined powder was pressed into cylindrical disks with a diameter of 13 mm by uniaxial pressing under 100 MPa. Finally, the samples were sintered at 1420 °C in air atmosphere for 2, 4 and 6h and designated BCZT-t, where t is the sintering dwell time.

### 2.3. Characterization equipments

The phase structure of the sintered ceramics was determined by X-ray diffraction (XRD, Panalytical X-Pert Pro) analysis using the Cu-Kα radiation with λ~1.540598 Å. The measurements were performed at room temperature in a range from 20° to 80° with the step size of 0.02°. Rietveld refinement analysis was achieved by using FullProf software.

The morphologies of the surfaces were revealed by a Scanning Electron Microscope (SEM, VEGA 3-Tscan) in association with energy dispersive X-ray (EDX) analysis. The densities of the pellets were measured using the Archimedes method [11]. The dielectric measurements and the impedance spectroscopy were examined by using an HP 4284A precision impedance meter, controlled by a computer in the 20Hz to 1MHz frequency range. The sintered samples surfaces were covered with a conductive silver paste serving as electrodes for electrical measurements. The polarization-electric field (P-E) hysteresis loops were collected using a commercial ferroelectric test system (TF Analyzer 3000, aixACCT). The electrocaloric adiabatic temperature variation (ΔT) and responsivity (ζ) were achieved by the indirect method from recorded (P-E) hysteresis loops at 1 Hz as a function of temperature.

## 3. Results and discussion

### 3.1. Structural analysis : XRD

**Figure.1** presents the X-ray diffraction patterns of BCZT ceramics sintered for various dwell times. All samples exhibit pure perovskite structure without any secondary phase in the limite of device detection, suggesting that $Ca^{2+}$ and $Zr^{4+}$ have diffused into the $BaTiO_3$ lattice to form a solid solution **[27]**. The Rietveld refinement of all ceramics shows the coexistence of tetragonal, T (space group P4mm) and orthorhombic, O (space group Amm2) phase at room temperature **[28], [29]**. Structural information and phase composition are enlisted in **Table. 1**. The crystallite size was determined using Scherrer's formula **[30]** as following :

$$D = \frac{k \times \lambda}{\beta \times cos\theta} \quad (1)$$

Where, k is the Scherrer constant, λ is the wavelength of the X-Rays used for the diffraction, β the « full width at half maximum » of the sharp peaks, and $\theta$ is the measured angle. From the **Table. 1**, it is observed that the crystallite size increases with increasing the dwell time, indicating better crystallinity for longer sintering time **[31]**.

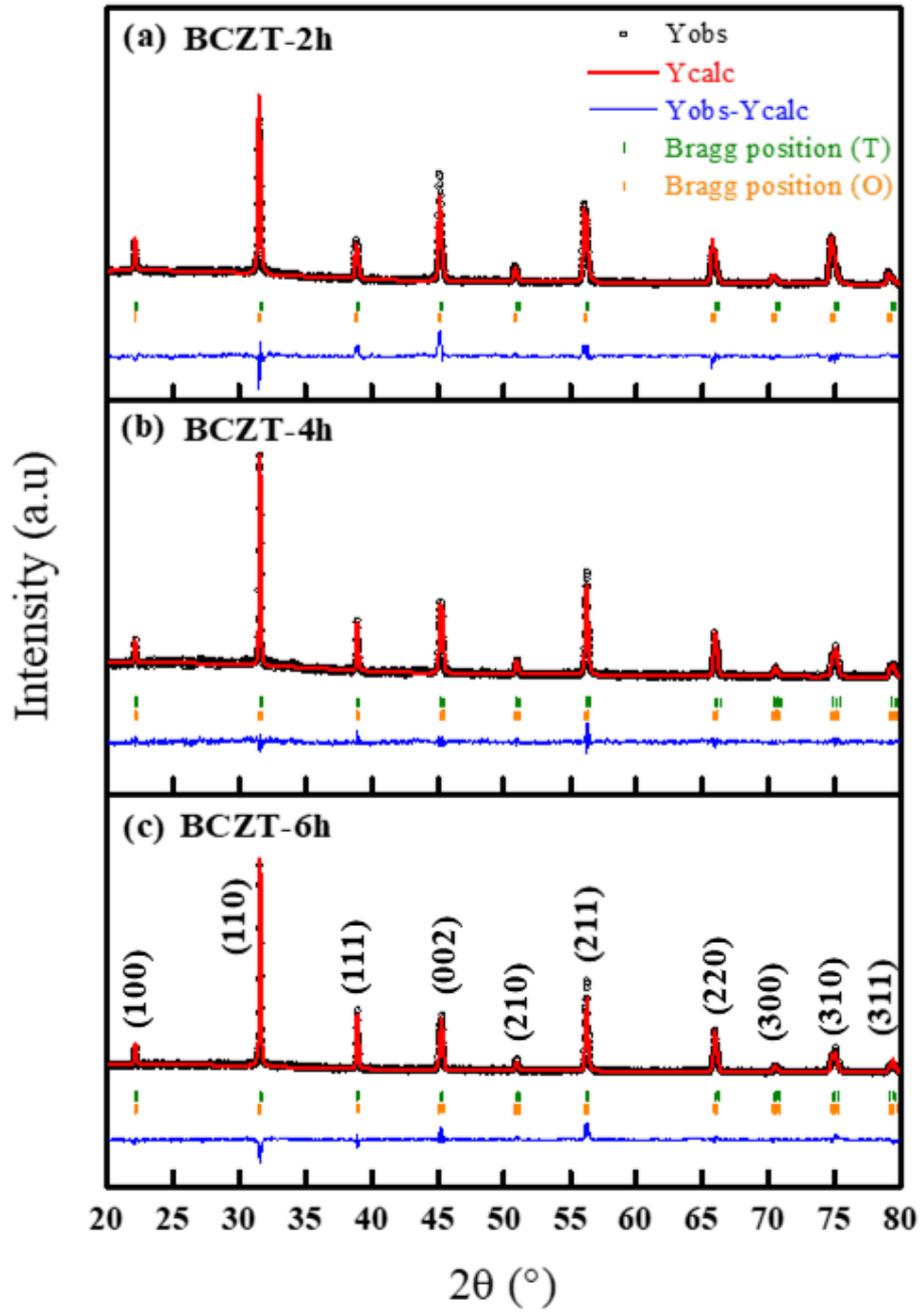

**Fig.1.** XRD patterns and Rietveld refinement of BCZT samples sintered for various dwell times : (a) 2h, (b) 4h and (c) 6h.

**Table. 1.** Structural parameters, average crystalline size, average grain size and relative density of BCZT ceramics sintered at different dwell times.

| Sintering time (h) | Structure | Unit cell parameters | | | | Tetragonality (c/a) | Phase compositions (wt%) | Reliability factors (%) | | | Average crystalline size (nm) by XRD | Average grain size(μm) by SEM | Relative density (%) |
|---|---|---|---|---|---|---|---|---|---|---|---|---|---|
| | | a(Å) | b(Å) | c(Å) | V (Å$^3$) | | | Rwp | Rp | $\chi^2$ | | | |
| 2 | Tetragonal + Orthorhombic | 3.9989 4.0188 | 3.9989 4.0130 | 4.0053 4.0144 | 64.052 64.744 | 1.0016 | 55.92 44.08 | 24.3 | 33.8 | 0.88 | 24.96 | 28.04 | 94.5 |
| 4 | Tetragonal + Orthorhombic | 3.9928 3.9957 | 3.9928 4.0134 | 4.0092 4.0053 | 63.918 64.231 | 1.0041 | 77.52 22.48 | 27.8 | 42.5 | 1.06 | 32.95 | 30.79 | 98.5 |
| 6 | Tetragonal + Orthorhombic | 3.9988 4.0162 | 3.9988 3.9952 | 4.0122 4.0073 | 64.159 64.300 | 1.0033 | 60.60 39.40 | 27.6 | 35.7 | 0.94 | 27.84 | 33.23 | 97.8 |

### 3.2. Microstructural and compositional analyses

**Figure. 2** displays the SEM images of the surface morphologies of the sintered BCZT ceramics with different sintering times of 2, 4 and 6h (insets show the corresponding histograms). It can be clearly seen that all the BCZT ceramics are densely sintered with irregular shaped and large grains up to micrometers. Furthermore, as shown in **Figure. 2a-c**. The average grain size estimated using an analytical software (image J) has increased from 28 to 33 μm with increasing sintering time from 2h to 6h. The relative density of all studied samples is greater than 94 %, this could be attributed to the highly active and well dispersion precursor powder prepared by the sol gel method **[32]**. Furthermore, it could be suggested that the sintering temperature of 1420°C was sufficient for the BCZT powder to produce good quality ceramics. The relevant grain size distributions and relative densities estimated are listed in **Table. 1**.

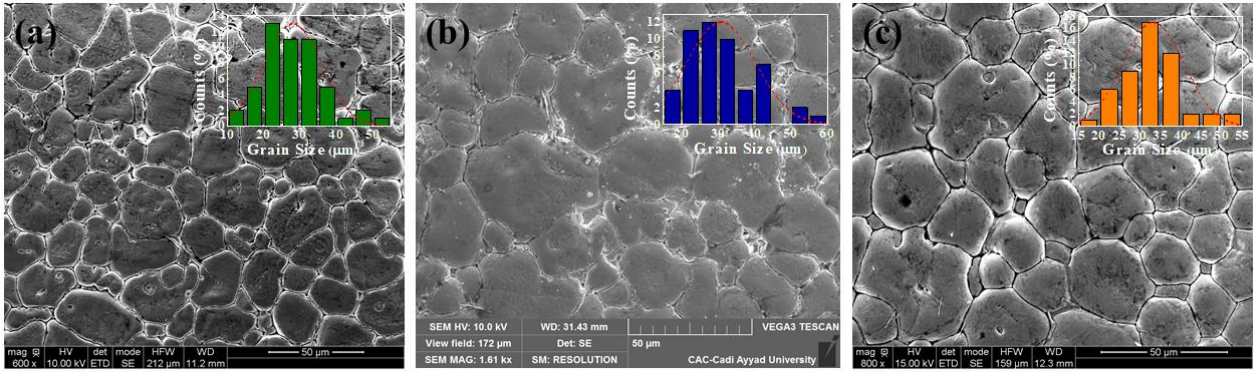

**Fig.2.** SEM micrographs and grain size distributions with Gaussian fitting (inset) of the samples sintered at various dwell times : (a) 2h, (b) 4h and (c) 6h.

To confirm the chemical composition of the BCZT ceramics, the ceramic elements distribution were performed by the EDX analysis, as presented in **Figure 3,** which indicates the presence of Ba, Ca, Ti, Zr and O elements. It also gives the ratios of Ba to Ca and Zr to Ti as 5.38 and 9.67 respectively, that confirms the as-prepared composition : as $Ba_{0.85}Ca_{0.15}Zr_{0.1}Ti_{0.9}O_3$. No other peak for any other element has been detected in the spectrum confirming the XRD observation and the chemical purity of the BCZT ceramics prepared by the sol-gel method.

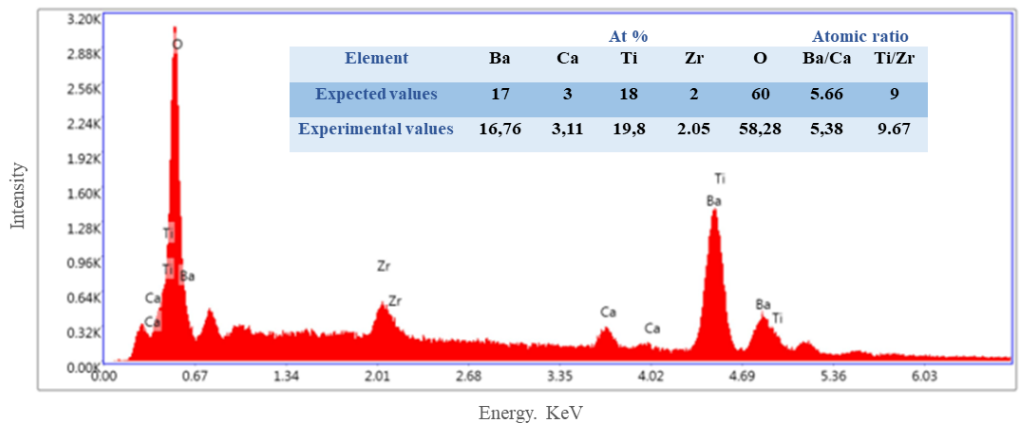

**Fig.3.** EDX spectra of BCZT ceramic sintered for 4h.

### 3.3. Dielectric analysis

The temperature dependence of the relative dielectric permittivity ($\varepsilon_r$) and dielectric loss factor (tan δ) of BCZT ceramics measured from room temperature to 200 °C are shown in **Figure. 4.** Two dielectric anomalies were observed for all samples: the first one corresponds to the morphotropic phase transition from the orthorhombic to tetragonal (O-T) phase, with increasing sintering time $T_{O-T}$ shifts from 31 to 24°C, while Tc shifts to higher temperature with increasing sintering time. The exact origin of this behavior is still unclear. The coexistence of the two phase structures in these ceramics may lead to good electrical properties. The dielectric permittivity maximum ($\varepsilon_{r,max}$) values depend greatly on sintering time as shown in **Table. 2**. It is worth noting that the BCZT-4h exhibits the largest value of dielectric permittivity ($\varepsilon_{r,max} \approx$ 16310, at 1 kHz), which is higher and/or similar to that obtained by other literature reported on BCZT ceramics prepared by different methods as illustrated in **Table. 3**. The significant improvement of the dielectric permittivity could be related to the high density of BCZT-4h **[33]**. All samples keep a low value of tan δ (≤ 0.06, at 1 kHz), this is probably attributed to the dense microstructure and the lower electron diffusion in the grain boundaries **[34]**. The overall properties (Tc, $\varepsilon_{r.max}$ $Tan\delta_{Tc}$) are recorded and summarized in **Table. 2**.

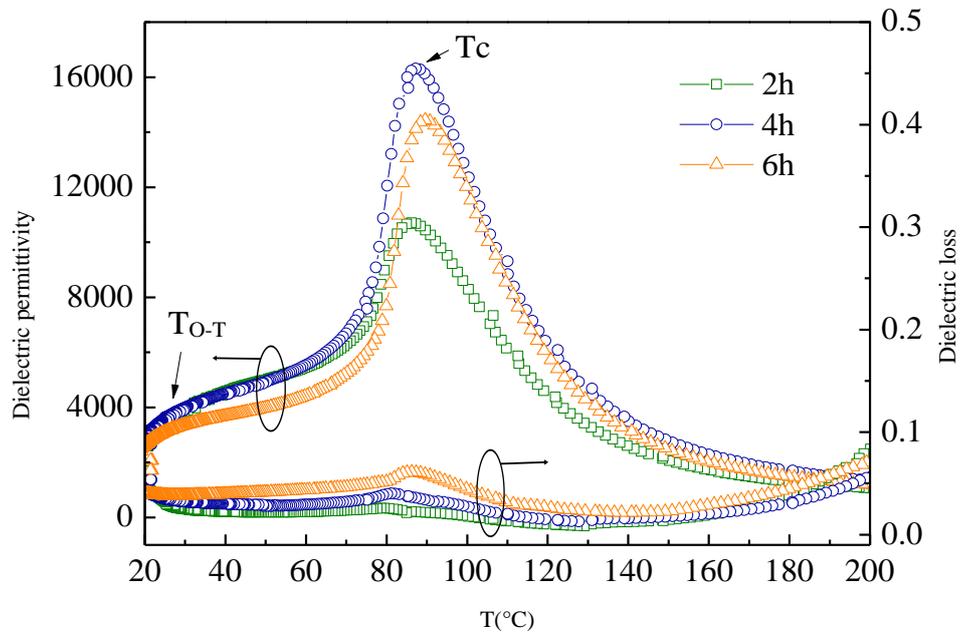

**Fig.4.** Variations of dielectric properties of BCZT samples sintered at various sintering times with the temperature measured at 1 kHz.

The temperature dependence of the dielectric permittivity for BCZT-4h ceramic at different frequencies, ranging from 100 Hz to 100 kHz is illustrated in **Figure. 5**. The first derivative of relative permittivity as a function of temperature is plotted (**inset of Figure. 5**). It is noticed that the dielectric maximum decreases as a function of frequency. This may be attributed to the decrease of the net polarization due to the accumulation of space charges at the grain boundaries which subsequently leads to the formation of interfacial polarization. In general, as frequency increases, the net polarization decreases as each polarization mechanism ceases to contribute, and hence, its dielectric constant decreases **[26], [35]**, however, there is no frequency dispersion in $T_C$. These two behaviors are associated with the diffuse phase transition in all samples and differs from the relaxor materials **[36]**.

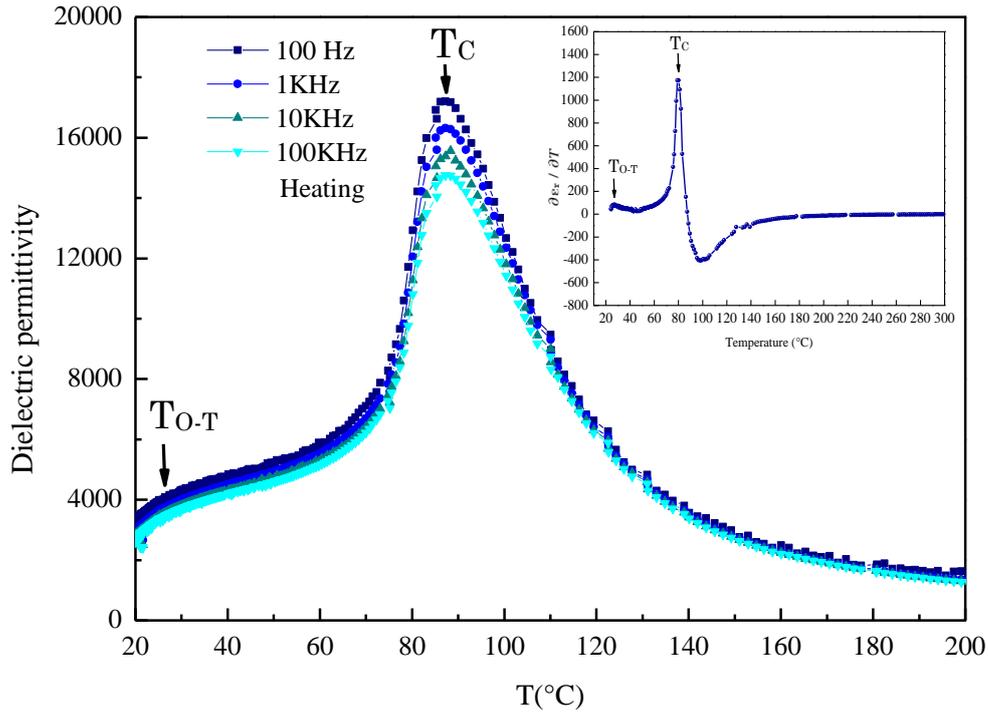

**Fig.5.** Temperature dependence of dielectric permittivity at different frequencies of BCZT samples sintered for 4h. Inset : Derivative of relative dielectric permittivity as a function of temperature.

It is known that the classical ferroelectrics follow the Curie-Weiss law above Curie temperature **[37]** which is defined by the following equation:

$$\frac{1}{\varepsilon_r} = \frac{T - T_0}{C} \quad (T > T_0) \qquad (2)$$

Where $T_0$ denotes the Curie–Weiss temperature and C is the Curie –Weiss constant reflecting the nature of the ferroelectrics transition.

The temperature-dependence of reciprocal dielectric permittivity at a frequency of 1kHz for BCZT-4h ceramic is presented in **Figure. 6**.

The values of the fitting results obtained by Eq. (2) are listed in the **Table. 2**. The Curie constant value for all samples is in the order of $10^5$ K, which are accordant with that of typical displacive-type ferroelectric **[38]**.

The changes of Curie–Weiss behavior ($\Delta T_m$) indicate the deviation from the Curie–Weiss law and the degree of diffuseness **[39]**, which can be defined by :

$$\Delta T_m = T_{cw} - T_m \qquad (3)$$

where $T_{cw}$ represents the temperature at which the dielectric permittivity begins to deviate from Curie-Weiss law, and $T_m$ is the temperature at maximum dielectric permittivity.

It is found that $\Delta T_m$ values are 39.74, 40.89 and 39.93°C for sintering times of 2, 4 and 6h, respectively, which implies a weaker diffuse phase transition behavior **[31]**.

To explain the diffuseness phase transition, a modified form of Curie-Weiss law **[40]**, proposed by Uchino is used which is given as follows:

$$\frac{1}{\varepsilon_r} - \frac{1}{\varepsilon_m} = \frac{(T - T_m)^\gamma}{C} \ (1 \leq \gamma \leq 2) \qquad (4)$$

where $\varepsilon_r$ is the relative permittivity, $\varepsilon_m$ is the maximum relative permittivity, $T_m$ is the temperature of the maximum dielectric permittivity, C is the Curie–Weiss constant and γ is the degree of diffuse phase transition, with 1< γ <2, this parameter gives information about the character of phase transition ; γ = 1 corresponds to normal ferroelectrics and γ = 2 corresponds to relaxor ferroelectrics. The inset in **Figure. 6** illustrates the linear fit plots of $\ln(1/\varepsilon_r - 1/\varepsilon_m)$ versus $\ln(T - T_m)$ at 1kHz for BCZT-4h ceramic and the value of γ is obtained from of the fitting curves. The values of γ are 1.760, 1.792 and 1.783 for 2, 4 and 6h, respectively, which indicate an incomplete diffuse phase transition behavior **[41]**.

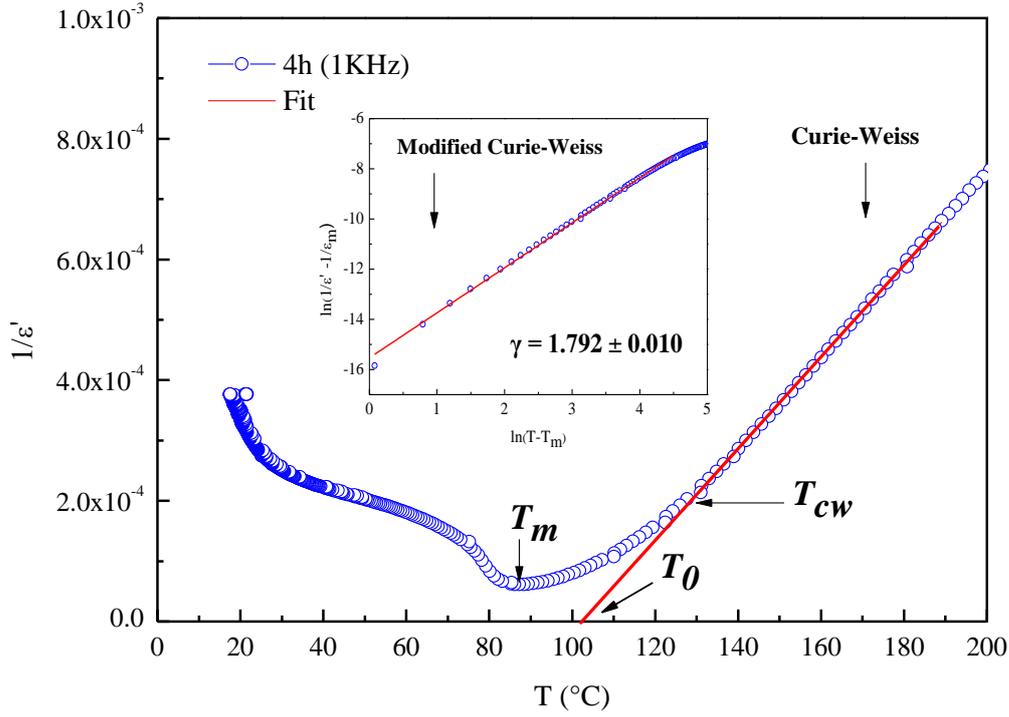

**Fig.6.** Curie–Weiss fitting curves of BCZT samples sintered for 4h.

**Table. 2.** Dielectric properties of BCZT ceramics performed at 1 kHz

| Sintering time (h) | $\varepsilon_{r,max}$ at 1kHz | $\tan\delta_{max}$ at 1kHz | $T_{O-T}$ (°C) | $T_m$ (°C) | $T_0$ (°C) | $T_{cw}$ (°C) | $\Delta T_m$ (°C) | $C \times 10^5$ (°C) | $\gamma$ |
|---|---|---|---|---|---|---|---|---|---|
| 2 | 10704 | 0,025 | 31 | 86.20 | 98.67 | 125.94 | 39.74 | 1.052 | 1.760 |
| 4 | 16310 | 0,040 | 27 | 87.22 | 102.61 | 128.11 | 40.89 | 1.302 | 1.792 |
| 6 | 14439 | 0,060 | 24 | 89.60 | 104.39 | 129.53 | 39.93 | 1.118 | 1.783 |

### 3.4. Ferroelectric analysis

In order to assess the EC effect, P-E hysteresis loops were registered for the BCZT-4h ceramic at different temperatures as showed in **Figure. 7**. It can be seen from the hysteresis loops that the increase of temperature increases the ferroelectric properties which is the typical characteristic of ferroelectric to paraelectric phase transition.

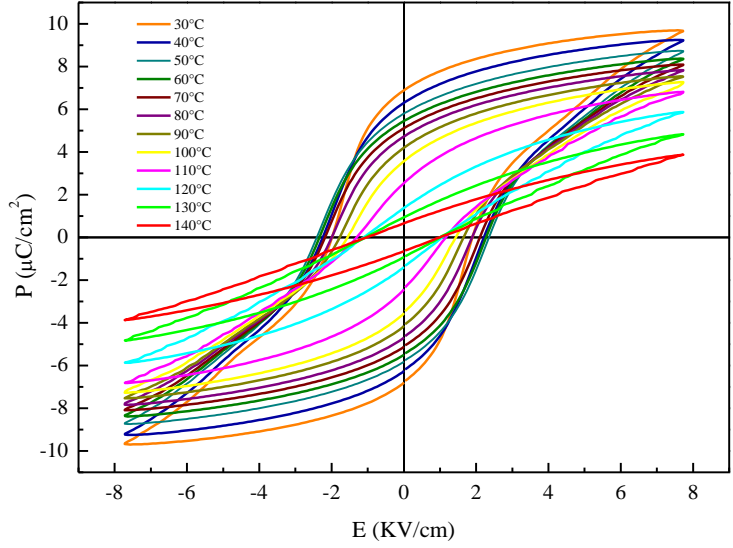

**Fig. 7.** P-E hysteresis loops of the BCZT samples sintered for 4h.

The grain size, density and phase homogeneity, affect the Ec, Ps, Pr and the squareness of the hysteresis loops. For an ideal hysteresis loop, the squareness parameter is equal to 2. According to Haertling and Zimmer, the squareness of the P-E loops can be quantitatively calculated using the following equation **[42], [43]** :

$$R_{sq} = \frac{P_r}{P_s} + \frac{P_{1.1Ec}}{P_r} \qquad (5)$$

Where, $R_{sq}$ is the squareness of hysteresis loop, Ps is saturation polarization, Pr is remnant polarization and $P_{1.1Ec}$ represents the polarization at an electric field equal to 1.1 times the coercive field (Ec). It can be seen that all samples exhibit approximately a value close to 2 (see **Table. 3**) indicating a better homogeneity and uniformity in grain size of the samples prepared by sol-gel wich contribute to a fast domain switching **[44]**.

**Figure. 8** displays the comparative (P-E) loops of polarization versus electric field of the sintered BCZT ceramics with various dwell time of 2, 4 and 6 h measured at room temperature under critical electric field. It can be clearly seen that the sintering dwell time has a significant effect on the ferroelectric properties (Ec, Ps, Pr) of BCZT ceramics. Indeed, the Ec of BCZT ceramics decreases from 2.3131 to 1.7427 kV/cm with an increase of dwell time from 2h to 6h. These values are lower and/or nearly the same as observed in others BCZT ceramics prepared by different processes and under various sintering conditions (see **Table. 3**). However, for BCZT-4h, Ps and Pr initially increase to a maximum values of 9.6955 and 6.8854 μC/cm$^2$ respectively and then decrease as showed in **Table. 3**. The great ferroelectric properties of BCZT-4h can be attributed to its high relative density (>98%) and large grain size (30.79µm), it could be understood that the BCZT-4h possesses the critical grain size over which the electrical properties can be enhanced.

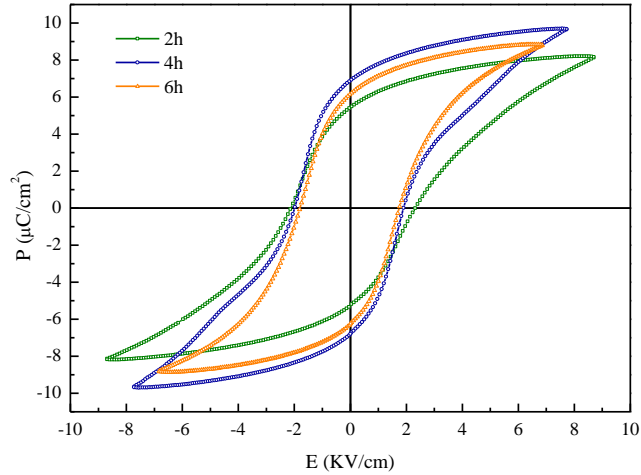

**Fig. 8.** Room-temperature P–E curves of the BCZT samples.

**Table. 3.** Comparaison of dielectric and ferroelectric properties of BCZT ceramics in the present work with others reported in literature for various synthesis metthods and sintering conditions.

| Sintering regime (°C-h) | Synthesis method | Average grain size(µm) | Relative density (%) | $\varepsilon_{r,max}$ | $\tan\delta_{max}$ | $R_{sq}$ | Ps (µC/cm$^2$) | Pr (µC/cm$^2$) | Ec (kV/cm) | E (kV/cm) | Refs |
|---|---|---|---|---|---|---|---|---|---|---|---|
| 1420-2 | Sol-gel | 28.04 | 94.5 | 10704 | 0,025 | 1.9653 | 8.2134 | 5.4410 | 2.3131 | 8.75 | This work |
| 1420-4 | Sol-gel | 30.79 | 98.5 | 16310 | 0,040 | 1.9295 | 9.6955 | 6.8854 | 1.8979 | 7.77 | This work |
| 1420-6 | Sol-gel | 33.23 | 97.8 | 14439 | 0,060 | 1.9424 | 8.8492 | 6.1749 | 1.7427 | 6.89 | This work |
| 1400-2 | Sol-gel | --- | 95 | 8808 | 0.02 | --- | 18 | 12.24 | 2.66 | 30 | [45] |
| 1420-6 | Sol-gel | --- | ---- | 16480 | 0.015 | --- | 17.76 | 11.64 | 1.78 | 30 | [19] |

| | | | | | | | | | | |
|---|---|---|---|---|---|---|---|---|---|---|
| 1350-2 | Sol-gel | 1.5 | --- | 6500 | 0.2 | --- | 3 | 0.7 | 1.9 | 12 | [46] |
| 1400-2 | Sol-gel-hydrothermal | --- | 95 | 9173 | --- | --- | 42 | 12.56 | ≈2.16 | 40 | [31] |
| 1300-3 | Hydrothermal | 12.09 | --- | 7760 | 0.1 | --- | 25 | 10.835 | 2.265 | 15 | [39] |
| 1500-4 | Solid state | 32.2 | 97.2 | 16300 | 0.017 | --- | 24 | 12 | 5 | 55 | [10] |
| 1500-2 | Solid state | 20.8 | --- | 9926 | 0.0108 | --- | --- | --- | --- | --- | [47] |
| 1300-3 | Solid state | 27.61 | 95 | 4500 | 0.12 | --- | 13.8 | 10.8 | 5.8 | 30 | [48] |

### 3.5. Energy storage performances

Generally, for nonlinear dielectrics, the recoverable energy density $W_{rec}$, total energy density $W_{total}$, and energy storage efficiency η can be determined from the polarization hysteresis (P-E) loops using the following equations **[49]**:

$$Wtotal = \int_{0}^{Pmax} EdP \qquad (6)$$

$$Wrec = \int_{Pr}^{Pmax} EdP \qquad (7)$$

and

$$\eta = \frac{Wrec}{Wtotal} \times 100 \qquad (8)$$

Where $P_{max}$, $P_r$, E, $W_{total}$, $W_{rec}$ and $\eta$ refer to maximum polarization, remnant polarization, applied external electric field strength, energy storage density, recoverable energy density, energy storage efficiency respectively. **Figure. 10** shows the variation of $W_{rec}$, $W_{loss}$ and η for BCZT ceramics as a function of sintering time. The BCZT-4h ceramic showed the better recoverable energy density (green area in **Figure. 9**) $W_{rec}$ (10.61 mJ/cm$^3$ at ~ 7.77 kV/cm), this improvement of $W_{rec}$ may be due to the large Ps values and low Ec values, which facilitate the reorientation of dipoles with less electrical field. The values of efficiency (η) of the samples are relatively comparable with others previously reported in the literature under strong electric field (see **Table. 4**). The best value of efficiency is $\eta=63.65\%$ at 140°C under an electric field of ~7.77 kV/cm, which is relatively comparable with others reported in the literature under strong electric field.

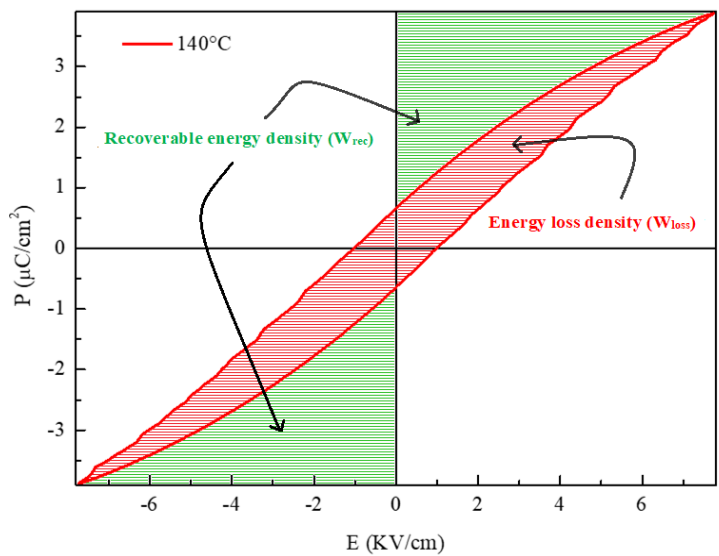

**Fig. 9.** P–E curve of the BCZT-4h at 140 °C with schematic for the calculation of energy storage performance.

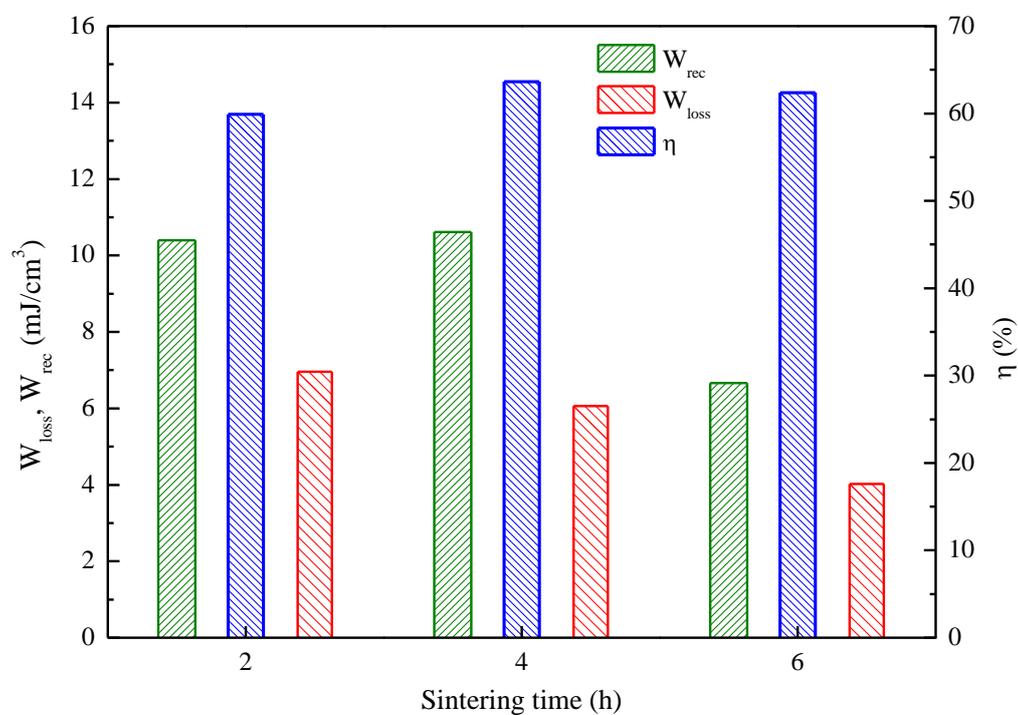

**Fig. 10.** Energy-storage properties of the BCZT samples with as a function of sintering time.

**Table. 4.** Comparison of energy storage properties of the BCZT family ceramics.

| Material | Sintering regime (°C-h) | Synthesis method | $W_{rec}$ (mJ/cm³) | $W_{loss}$ (mJ/cm³) | $\eta$ (%) | E (kV/cm) | Refs |
|---|---|---|---|---|---|---|---|
| $Ba_{0.85}Ca_{0.15}Zr_{0.1}Ti_{0.9}O_3$ | 1420-2 | Sol-gel | 10.40 | 6.96 | 59.90 | 8.75 | This work |
| $Ba_{0.85}Ca_{0.15}Zr_{0.1}Ti_{0.9}O_3$ | 1420-4 | Sol-gel | 10.61 | 6.06 | 63.65 | 7.77 | This work |
| $Ba_{0.85}Ca_{0.15}Zr_{0.1}Ti_{0.9}O_3$ | 1420-6 | Sol-gel | 6.66 | 4.02 | 62.36 | 6.89 | This work |
| $Ba_{0.85}Ca_{0.15}Zr_{0.1}Ti_{0.9}O_3$ | 1420-6 | Sol-gel | 520 | --- | --- | 30 | [19] |
| $Ba_{0.85}Ca_{0.15}Zr_{0.1}Ti_{0.9}O_3$ | 1420-6 | Solid state | 310 | --- | --- | 30 | [19] |
| $[(BaZr_{0.2}Ti_{0.80})O_3]_{0.85}[(Ba_{0.70}Ca_{0.30})TiO_3]_{0.15}$ | 1600-4 | Solid state | 940 | 680 | 72 | 170 | [50] |
| $Ba_{0.85}Ca_{0.15}Zr_{0.1}Ti_{0.9}O_3$ | 1320-6 | Co-precipitation | 250 | 130 | 65 | 100 | [51] |
| $Ba_{0.95}Ca_{0.05}Zr_{0.3}Ti_{0.7}O_3$ | 1280-2 | Citrate method | 590 | --- | 72.8 | 160 | [52] |

### 3.6. Indirect electrocaloric measurements

Using Maxwell relation $(\frac{\partial P}{\partial T})_E = (\frac{\partial S}{\partial E})_T$ the adiabatic temperature change ΔT under an applied electric field is calculated by the following equation **[53]**:

$$\Delta T = -\frac{1}{\rho}\int_{E2}^{E1} \frac{T}{Cp}(\frac{\partial P}{\partial T})_E \, dE \qquad (9)$$

where ρ is the density of the ceramics, E1 and E2 are the lower and higher electric field limits, respectively, Cp denotes the specific heat capacity of the ceramics.

There are several factors that can influence the performance of the electrocaloric materials, the nature of the phase transition is one of these factors. Two types of transitions can be distinguished; first order transition is associated with a high value of ΔT with a restricted temperature range around Tc, while the second order transition is generally characterized by a low electrocaloric effect with a large temperature range around Tc **[54]**.

The electrocaloric (EC) properties of the fabricated BCZT ceramics for an applied electric field 6 kV/cm are shown in **Figure. 11**. The BCZT-4h and BCZT-6h present high values of $\Delta T_{EC,max}$ (0.1467K and 0.1513K respectively) in a limited temperature ranges, this may be related to the increase of the spontaneous polarization. This behavior indicates that the phase transition is of first order. However, BCZT-2h presents very low value of $\Delta T_{EC,max}$ with relatively large temperature range (FWHM$_{\Delta T}$) of ≈ 56K, which indicates that the transition is of second order.

The EC responsivity ($\xi=\Delta T/\Delta E$) values obtained for the BCZT samples are enlisted in **Table. 5**, comparing with different studies reported in literature. We notice that the BCZT ceramics sintered at 4h and 6h exhibit high EC responsivity of 0.244 K mm/kV and 0.252 K mm/kV respectively at low applied electric field of 6 kV/cm. These values appear to be comparable and/or higher than those of the BCZT ceramics prepared by solid state and sol-gel routes (see **Table. 5**). It is worth mentioning that the most important criteria that have to be considered during development of EC materials is the fact that the EC responsivity ($\Delta T/\Delta E$) should be as high as possible within a large interval of temperature **[55]**. The BCZT-4h exhibits a better EC responsivity with relatively large temperature interval (FWHM$_{\Delta T}$) of ≈ 33K.

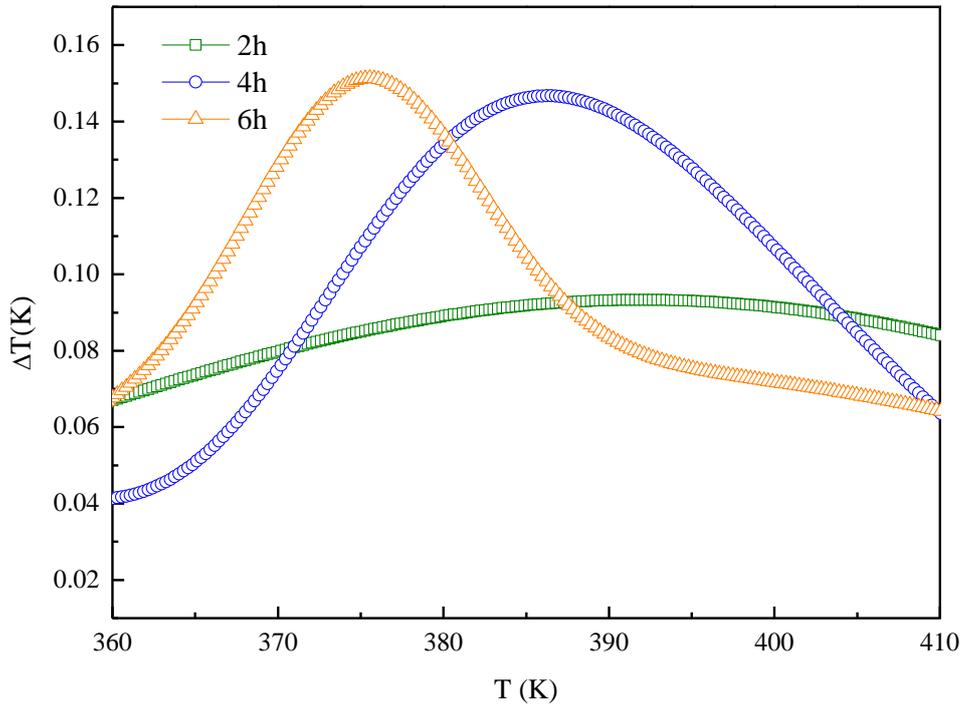

**Fig. 11.** Temperature profiles of the electrocaloric temperature change ($\Delta T$) for various sintering times at 6 kV/cm applied electric field.

**Table. 5.** Comparison of the ECE properties of BCZT ceramics with other lead-free materials reported in literature.

| Material | Sintering regime (°C-h) | FWHM$_{\Delta T}$ (K) | $\Delta T_{EC,max}$ (K) | $\Delta E$ (kV/cm) | $\xi$ (K mm/kV) | Synthesis method | Measurement method | Refs |
|---|---|---|---|---|---|---|---|---|
| Ba$_{0.85}$Ca$_{0.15}$Zr$_{0.1}$Ti$_{0.9}$O$_3$-2h | 1420-2 | 56 | 0.0933 | 6 | 0.155 | Sol-gel | Indirect | Present work |
| Ba$_{0.85}$Ca$_{0.15}$Zr$_{0.1}$Ti$_{0.9}$O$_3$-4h | 1420-4 | 33 | 0.1467 | 6 | 0.244 | Sol-gel | Indirect | Present work |
| Ba$_{0.85}$Ca$_{0.15}$Zr$_{0.1}$Ti$_{0.9}$O$_3$-6h | 1420-6 | 23 | 0.1513 | 6 | 0.252 | Sol-gel | Indirect | Present work |
| Ba$_{0.85}$Ca$_{0.15}$Zr$_{0.1}$Ti$_{0.9}$O$_3$ | 1350-2 | --- | 0.0150 | 6 | 0,025 | Sol-gel | Indirect | [46] |
| Ba$_{0.85}$Ca$_{0.15}$Zr$_{0.1}$Ti$_{0.9}$O$_3$ | 1420-2 | --- | 0.1520 | 8 | 0.190 | Solid state | Indirect | [56] |
| Ba$_{0.8}$Ca$_{0.2}$Zr$_{0.06}$Ti$_{0.94}$O$_3$ | 1420-2 | --- | 0.2100 | 7.95 | 0.264 | Solid state | Indirect | [57] |
| 0.65Ba(Zr$_{0.2}$Ti$_{0.8}$)O$_3$-0.35(Ba$_{0.7}$Ca$_{0.3}$)TiO$_3$ | 1500-2 | --- | 0.3300 | 20 | 0.165 | Solid state | Direct | [58] |
| Ba$_{0.98}$Ca$_{0.02}$(Zr$_{0.085}$Ti$_{0.915}$)O$_3$ | 1280-2 | --- | 0.6000 | 40 | 0.150 | Solid state | Direct | [59] |
| Ba$_{0.85}$Ca$_{0.15}$Zr$_{0.1}$Ti$_{0.9}$O$_3$ | 1400-4 | --- | 0.4000 | 21.5 | 0.186 | Solid state | Indirect | [60] |
| Ba$_{0.8}$Ca$_{0.2}$Zr$_{0.04}$Ti$_{0.96}$O$_3$ | 1350-2 | --- | 0.2700 | 7.95 | 0.340 | Solid state | Indirect | [61] |

## 4. Conclusion

In this contribution, we have dealt with the effect of sintering time on the structural, dielectric, ferroelectric and piezoelectric properties of lead-free Ba$_{0.85}$Ca$_{0.15}$Ti$_{0.9}$Zr$_{0.1}$O$_3$ (BCZT) ceramics prepared by sol-gel route. BCZT ceramic fabricated at 1420 °C for 4h with relative density of 98% has demonstrated simultaneously enhanced dielectric properties around the MPB region, i.e, $\varepsilon_{r.max}$ = 16310 and improved electrocaloric effect; $\zeta$ = 0.244 K mm/kV. In addition, BCZT-4h displayed a relatively high Wrec of 10.61 mJ/cm$^3$ with an efficiency coefficient of ~ 63% at low electric field strength. These results suggest that the synthesized BCZT-4h ceramic could be a promising candidate for electrocaloric and energy storage applications.

**Acknowledgements**

The authors gratefully acknowledge the financial support of the European H2020-MSCA-RISE-2017-ENGIMA action.
**References**

[1] Z. Sun *et al.*, « Dielectric and piezoelectric properties and PTC behavior of Ba0.9Ca0.1Ti0.9Zr0.1O3−xLa ceramics prepared by hydrothermal method », *Mater. Lett.*, vol. 118, p. 1-4, mars 2014, doi: 10.1016/j.matlet.2013.12.043.

[2] R. Bhimireddi, B. Ponraj, et K. B. R. Varma, « Structural, Optical, and Piezoelectric Response of Lead-Free $Ba_{0.95}Mg_{0.05}Zr_{0.1}Ti_{0.9}O_3$ Nanocrystalline Powder », *J. Am. Ceram. Soc.*, vol. 99, nº 3, p. 896-904, mars 2016, doi: 10.1111/jace.14018.

[3] S. Saparjya, T. Badapanda, S. Behera, B. Behera, et P. R. Das, « Effect of Gadolinium on the structural and dielectric properties of BCZT ceramics », *Phase Transit.*, vol. 93, nº 2, p. 245-262, févr. 2020, doi: 10.1080/01411594.2020.1711905.

[4] S. D. Chavan et D. J. Salunkhe, « Ferroelectric Relaxation Behavior of Lead free BCZT Ceramics », vol. 3, nº 11, p. 5, 2012.

[5] J. Tao, Z. Yi, Y. Liu, M. Zhang, et J. Zhai, « Dielectric Tunability, Dielectric Relaxation, and Impedance Spectroscopic Studies on $(Ba_{0.85}Ca_{0.15})(Ti_{0.9}Zr_{0.1})O_3$ Lead-Free Ceramics », *J. Am. Ceram. Soc.*, vol. 96, nº 6, p. 1847-1851, juin 2013, doi: 10.1111/jace.12265.

[6] S. Ben Moumen *et al.*, « Impedance spectroscopy studies on lead free $Ba_{1-x}Mg_x(Ti_{0.9}Zr_{0.1})O_3$ ceramics », *Superlattices Microstruct.*, vol. 118, p. 45-54, juin 2018, doi: 10.1016/j.spmi.2018.04.012.

[7] W. Cai *et al.*, « Synergistic effect of grain size and phase boundary on energy storage performance and electric properties of BCZT ceramics », *J. Mater. Sci. Mater. Electron.*, avr. 2020, doi: 10.1007/s10854-020-03446-z.

[8] S. Ben Moumen *et al.*, « Structural, Dielectric, and Magnetic Properties of Multiferroic $(1-x)La_{0.5}Ca_{0.5}MnO_3$-$(x)BaTi_{0.8}Sn_{0.2}O_3$ Laminated Composites », *IEEE Trans. Ultrason. Ferroelectr. Freq. Control*, vol. 66, nº 12, p. 1935-1941, déc. 2019, doi: 10.1109/TUFFC.2019.2935459.

[9] P. Jaimeewong, M. Promsawat, A. Watcharapasorn, et S. Jiansirisomboon, « Comparative study of properties of BCZT ceramics prepared from conventional and sol-gel auto
19